\documentclass[12pt,twoside]{article} 
\usepackage{amsfonts}
\usepackage{graphicx} 

\newtheorem{theorem}{Theorem}[section]
\newtheorem{lemma}[theorem]{Lemma}
\newtheorem{proposition}[theorem]{Proposition}

\newtheorem{rmrk}[theorem]{Remark}

\newcommand{\R} {{\mathbb R}}

\newcommand{\qed} {\hfill {\small Q.E.D.} \par\medskip}
\newcommand{\skippar} {\par\medskip}
\newcommand{\ds} {\displaystyle}
\newcommand{\proof} {\noindent \textsc{Proof.} }
\newcommand{\proofof}[1] {\noindent \textsc{Proof of {#1}.} }
\newcommand{\article}[3] {\textsc{{#1}}, {\itshape {#2}}, {{#3}}.}
\newcommand{\book}[3] {\textsc{{#1}}, {\itshape {#2}}, {{#3}}.}
\newcommand{\vol} {\textbf}
\newcommand{\eps} {\varepsilon}
\newcommand{\rset}[2] {\left\{ #1 \: \left| \: #2 \right. \! \right\} }

\renewcommand{\tilde} {\widetilde}
\renewcommand{\o} {orbit}

\newcommand{\fn} {function}
\newcommand{\bi} {billiard}
\newcommand{\me} {measure}
\newcommand{\tr} {trajector}

\newcommand{\erg} {ergodic}
\newcommand{\ex} {existence}

\newcommand{\sy} {system}
\newcommand{\hyp} {hyperbolic}
\newcommand{\ta} {Q}		
\newcommand{\f} {f}		
\newcommand{\ph} {\varphi}	
\newcommand{\si} {\mathcal{S}}	
\newcommand{\z} {z}		
\newcommand{\w} {w}		
\newcommand{\uw} {\mathcal{U}}	
\newcommand{\ma} {T}		
\newcommand{\ps} {\mathcal{M}}	

\renewcommand{\r} {r}		

\newcommand{\sect}[1] {\section{{#1}} \setcounter{equation}{0}}

\newcommand{\fig}[3] {
\medskip\smallskip
\begin{figure}[ht]
	\centering
	\includegraphics[width=#2]{#1.eps}
	\begin{minipage}[t]{0.75\linewidth} 
		\caption{\baselineskip=14pt {#3}}
		\protect\label{#1}
	\end{minipage}
\end{figure}
\medskip
}

\newenvironment{remark}
{\begin{rmrk} \em}
{\end{rmrk}}

\begin{document}

\title{More ergodic billiards with an infinite cusp}

\author{
	Marco Lenci \\
	Department of Mathematical Sciences \\
	Stevens Institute of Technology \\
	Hoboken, NJ \ 07030 \\ 
	U.S.A. \\
	\footnotesize{E-mail: \texttt{mlenci@stevens-tech.edu}}
}

\date{January 2002}

\maketitle

\begin{abstract}
	In \cite{le2} the following class of billiards was studied: For
	$f: [0, +\infty) \longrightarrow (0, +\infty)$ convex,
	sufficiently smooth, and vanishing at infinity, let the
	billiard table be defined by $Q$, the planar domain delimited
	by the positive $x$-semiaxis, the positive $y$-semiaxis, and
	the graph of $f$.

	For a large class of $f$ we proved that the billiard map was
	hyperbolic. Furthermore we gave an example of a family of $f$
	that makes this map ergodic. Here we extend the latter result
	to a much wider class of functions.

\end{abstract}

\sect{Introduction}
\label{sec-intro}

This note is a follow-up to \cite{le2}, where we studied a certain
family of \bi s in the plane. A \bi\ is a dynamical \sy\ defined by
the free motion of a material point inside a domain, called the
\emph{table}, with the prescription that when the point hits the
boundary of the table it gets reflected at an angle equal to the angle
of incidence.

The family introduced in \cite{le2} was defined as follows: To a
three-times differentiable \fn\ $\f: [0, +\infty) \longrightarrow (0,
+\infty)$, convex, vanishing at $+\infty$ (and thus bounded), we
associate the table $\ta$ delimited by the positive $x$-semiaxis, the
positive $y$-semiaxis, and the graph of $f$, as in Fig.~\ref{figa}.

\fig{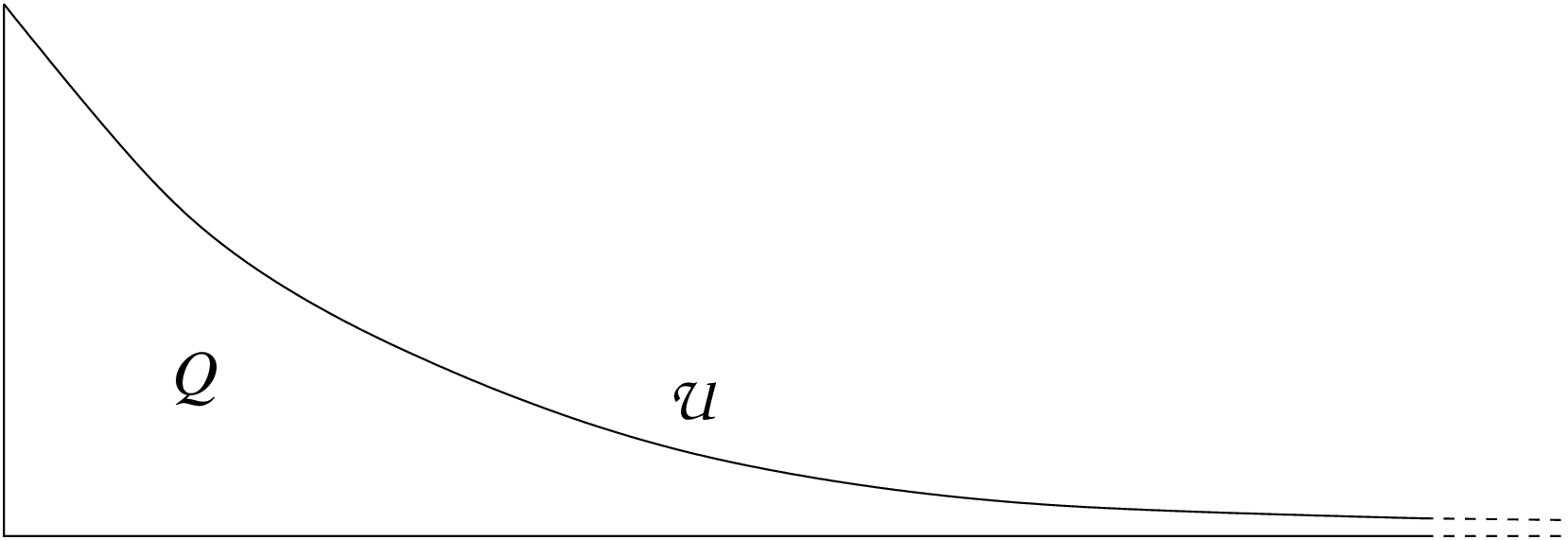} {4in} {An example of a \bi\ table $\ta$.} 

These \bi s are \emph{semi-dispersing}, which means that the
boundaries of the tables are made of flat and convex parts, when
seen from the inside. The main feature of these domains, however, is
that they are unbounded; in particular they possess a non-compact
\emph{cusp}.

There are several reasons why models of the like merit study. From a
physical point of view they are on the borderline between
\emph{closed} and \emph{open} \sy s. By definition, a closed \sy\ is
one in which the evolution of any representative point in the relevant
phase space remains confined within a compact set. An open \sy, on the
contrary, possesses unbounded \o s.  Our case is borderline in the
sense that, albeit open, an arbitrarily small perturbation of the \sy\
will make it closed. Some suggest these \sy s be called \emph{ajar}.

In general it is reasonable to expect ajar \sy s to have many
unbounded \o s, in the sense of a positive-\me\ set of initial
conditions, with respect to a suitable \me. Moreover, if the \sy\
enjoys some stochastic properties (e.g., \erg ity) one might expect
that almost all \o s are unbounded. What might really discriminate
between substantially different dynamics for two distinct ajar \sy s
is the number of \emph{escape \o s}. In a few words, an escape \o\ is
a \tr y that eventually leaves for good any given compact set of the
phase space.

As an example of two opposite behaviors, consider a particle in a
Newtonian potential with total energy zero. We know that all \tr ies,
apart from the singular ones that intersect the center of the
potential, are parabolas, hence escape both in the past and in the
future. On the other hand, the \bi\ in Fig.~\ref{figa} has only one
escape \o---the obvious one that runs along the $x$-semiaxis \cite{l,
k, le1}.

From a more mathematical standpoint the relevance of our \bi s lies in
the fact that they give rise to infinite-\me\ \hyp\ dynamical \sy s,
about which not much is known. We do not delve in this issue here, as
a good part of the introduction of \cite{le2} is devoted to it. Let us
just mention that, if one is interested in studying the stochastic
properties of these \sy s (e.g., \erg ity, as in our case), one cannot
apply the fundamental results of \cite{ks}, where a version of Pesin's
theory is engineered to be used on \emph{finite-\me} \hyp\ \sy s with
singularities. Other standard results that are denied to us include
the estimatates on the \me\ of the tubular neighborhoods of the
singularities and, what's more, the local ergodicity theorem (also
known as \emph{fundamental theorem}).

\skippar

The content of this paper is well summarized by its title: we are
presenting a (fairly large) class of \erg\ \bi s like the above. In
\cite{le2} we expounded the underpinning ideas and technical
arguments, but produced only a one-parameter family of
examples---those defined by $\f(x) = C x^{-p}$, with $C,p>0$. Here we
extend the result to general perturbations of the above family. To do
so, we need to assume on the reader's side a rather strong grasp on
\cite{le2} (which in turn is heavily based on the seminal article
\cite{lw}).

After some generalities, outlined in Section \ref{sec-gen}, we will
verify that a large class of \fn s $\f$ satisfies all the general
theorems of \cite{le2}, the ones establishing hyperbolicity; this is
done in Section \ref{sec-hyp}. In Section \ref{sec-erg} we will work
on \erg ity. In particular, we will present a lemma that substitutes a
corresponding lemma of \cite{le2} and extends its conclusions to our
new class of \fn s.

\skippar

Before moving on to the mathematics proper, it might be worthwhile to
recall what we mean by \erg ity here, as this might lend itself to
ambiguities in the case of phase spaces of infinite \me. We say that a
dynamical \sy, endowed with an invariant \me\ $\mu$, is \erg\ when the
time average of every integrable \fn\ is constant almost everywhere
w.r.t.\ $\mu$ \cite[Defn.~8.1]{le2}. Although this is a rather weak
notion of \erg ity (it does not even prevent the \ex\ of two
complementary invariant subsets of infinite \me) we also have a much
more satisfactory result: the \erg ity of the (finite-\me) Poincar\'e
map corresponding to the returns onto the vertical side of the
boundary \cite[Prop.~8.11]{le2}.

\sect{Main result and generalities}
\label{sec-gen}

In the rest of this note we show that the class of functions defined
hereafter yields \erg\ \bi s, in the sense mentioned above (see also
\cite[Sec.~8]{le2}).

Assume that $\f: [0, +\infty) \longrightarrow (0, +\infty)$ is three
times differentiable, convex, and vanishing at $+\infty$; and
that there exists a \fn\ $\upsilon$ such that
$$
	\f(x) = \upsilon(x) \, x^{-p}, \qquad \mbox{for some } p>0;
	\eqno{\mathrm{(H1)}}
$$
$$
	\upsilon(x) \, x^{-\eps} \to 0, \qquad \mbox{for every } \eps>0;
	\eqno{\mathrm{(H2)}}
$$
$$
	\frac{\upsilon^{(i)}(x) x^i} {\upsilon(x)} \to 0, \qquad
	i=1,2,3.  
	\eqno{\mathrm{(H3)}}
$$
The limit in question is always $x \to +\infty$. In (H3),
$\upsilon^{(i)}$ denotes the $i$-th derivative of $\upsilon$. 

In practical terms, this means that $\f(x)$ is a well-behaved
perturbation of $x^{-p}$, even at the level of a few derivatives (see
Lemma \ref{lemma-simeq}). The family $\f(x) = C x^{-p}$, $C>0$, was
shown in \cite{le2} to give \erg\ tables and is here superseded by
this wider class. That this family is much more general is shown by the
next few examples, that are easily checked against (H1)-(H3).

\begin{itemize}

\item $\f(x) = \log^\gamma x \, x^{-p}$, with $\gamma \in \R$; thus
$\upsilon(x) = \log^\gamma x$.

\item $\f(x) = a x^{-p} + b x^{-q}$, with $0<p<q$, $a>0$, $b \in
\R$; take $\upsilon(x) = a + b x^{-q+p}$.

\item $\ds \f(x) = \frac1 {a x^p + b x^q}$, with $p>q>0$, $a>0$, $b \in
\R$; use $\upsilon(x) = \ds \frac1 {a + b x^{-p+q}}$.

\end{itemize}

\begin{remark}
	Not all \fn s listed above are necessarily positive and convex
	for $x \ge 0$, but certainly they become so
	asymptotically. The reader is therefore invited to think of
	them as shifted to the left as much as is needed.
\end{remark}

The second of the above examples is actually an instance of a more
general result about the family at hand:

\begin{proposition}
	The space of sufficiently smooth, positive, convex, vanishing
	\fn s that verify \emph{(H1)-(H3)} is positive-linear; that
	is, a linear combination of elements in the space, with
	positive coefficients, belongs to the space.
	\label{prop-lin}
\end{proposition}

\proof That the space in question is homogeneous is obvious. We only
need prove that it is additive. Therefore, for $j=1,2$, take $\f_j(x)
= \upsilon_j(x) \, x^{-p_j}$. To fix the ideas, suppose that $p_1 <
p_2$. (H3) amounts to saying that, for $i=1,2,3$ and $j=1,2$, there
exist six positive \fn s $g_j^{(i)}$ such that
\begin{equation}
	\left| \upsilon_j^{(i)}(x) \right| x^i = g_j^{(i)}(x)
	\upsilon_j(x)
\end{equation}
and $g_j^{(i)}(x) \to 0$, as $x \to +\infty$. In order to rewrite $f :=
f_1 + f_2$ in terms of (H1), we choose $p := p_1$ and
\begin{equation}
	\upsilon(x) := \upsilon_1(x) + \upsilon_2(x) \, x^{-\alpha},
\end{equation}
with $\alpha := p_2 - p_1 > 0$. Evidently $\upsilon$ verifies (H2). We
show that it verifies (H3) as well. Let us do it only for $i=2$, the
other cases just being analogous. Since
\begin{equation}
	\upsilon''(x) = \upsilon_1''(x) + \upsilon_2''(x) x^{-\alpha}
	- 2 \alpha \upsilon_2'(x) x^{-\alpha-1} + \alpha (\alpha+1)
	\upsilon_2(x) x^{-\alpha-2},
\end{equation}
then
\begin{eqnarray}
	|\upsilon''(x)| x^2 &\le& |\upsilon_1''(x)| x^2 +
	|\upsilon_2''(x)| x^2 x^{-\alpha} + 2 \alpha |\upsilon_2'(x)|
	x x^{-\alpha} + \alpha (\alpha+1) \upsilon_2(x) x^{-\alpha}
	\nonumber \\
	&\le& g_1^{(2)}(x) \upsilon_1(x) + g_2^{(2)}(x) \upsilon_2(x)
	x^{-\alpha} + \\
	&& \qquad + 2 \alpha g_2^{(1)}(x) \upsilon_2(x) x^{-\alpha}
	+ \alpha (\alpha+1) \upsilon_2(x) x^{-\alpha} \nonumber \\
	&\le& h(x) \left[ \upsilon_1(x) +  \upsilon_2(x) \right], 
	\nonumber
\end{eqnarray}
with
\begin{equation}
	h(x) := 3 \max \left\{ g_1^{(2)}(x), g_2^{(2)}(x) x^{-\alpha},
	2 \alpha g_2^{(1)}(x) x^{-\alpha}, \alpha (\alpha+1)
	x^{-\alpha} \right\}.
\end{equation}
But $h(x) \to 0$, as $x \to +\infty$; which gives (H3) when $i=2$.  
\qed

The traditional way to study a \bi\ problem is to look at the \sy\
only when the point collides against the boundary of the
table. Mathematically speaking, this means that we consider the
Poincar\'e map defined by the \bi\ flow on the cross-section that
comprises all vectors of the unit tangent bundle of $\partial \ta$
that point inside $\ta$ (i.e., immediately after the collision). The
restriction to unit vectors is obvious as the original Hamiltonian
\sy\ preserves the modulus of the velocity, which is conventionally
fixed to 1.  These vectors are traditionally called \emph{line
elements} \cite{s,cfs}.

In our case, the geometry of the table suggests that we further
restrict the cross-section to vectors based in $\uw$, the curved
(dispersing) part of the boundary, as indicated in Fig.~\ref{figa}. In
fact, setting $\ta_4 := \rset{ (x,y) \in \R \times \R } { |y| \le
\f(|x|) \, }$ (see Fig.~\ref{fige} later on), one can think of working
with the \bi\ in $\ta_4$ where the four smooth components of $\partial
\ta_4$ (together with all unit vectors thereupon based) have been
identified. This is called a four-copy \emph{unfolding} of $\ta$.

Line elements can be parametrized by $\z := (\r, \ph) \in \ps :=
(0,+\infty) \times (0,\pi)$, where $\r$, the arc-length along $\uw$,
specifies the base point of the vector, and $\ph$ is the angle that
this vector makes with the tangent to $\uw$ there. We choose the
convention that $\ph$ close to zero denotes almost tangential vectors
that point towards the infinite cusp. The \bi\ map $\ma$ is defined at
all points of $\ps$ that would not end up in a vertex or hit $\partial
\ta_4$ tangentially; that is, we morally exclude the set ``$\ma^{-1}
\partial \ps$''. In fact, these points make up the discontinuity set
of $\ma$---commonly called the singularity set, as $\ma$ there is more
often singular than not. It is easy to see that this set, denoted
$\si^{+}$, is the union of two curves, $\si^{1+}$ and $\si^{2+}$,
depicted in Fig.~\ref{figd}. To conclude this hasty description of the
dynamical \sy\ at hand, we recall that $\ma$ preserves the \me\
$d\mu(\r,\ph) = \sin\ph \, d\r d\ph$. More details are available in
\cite{le2}.

\fig{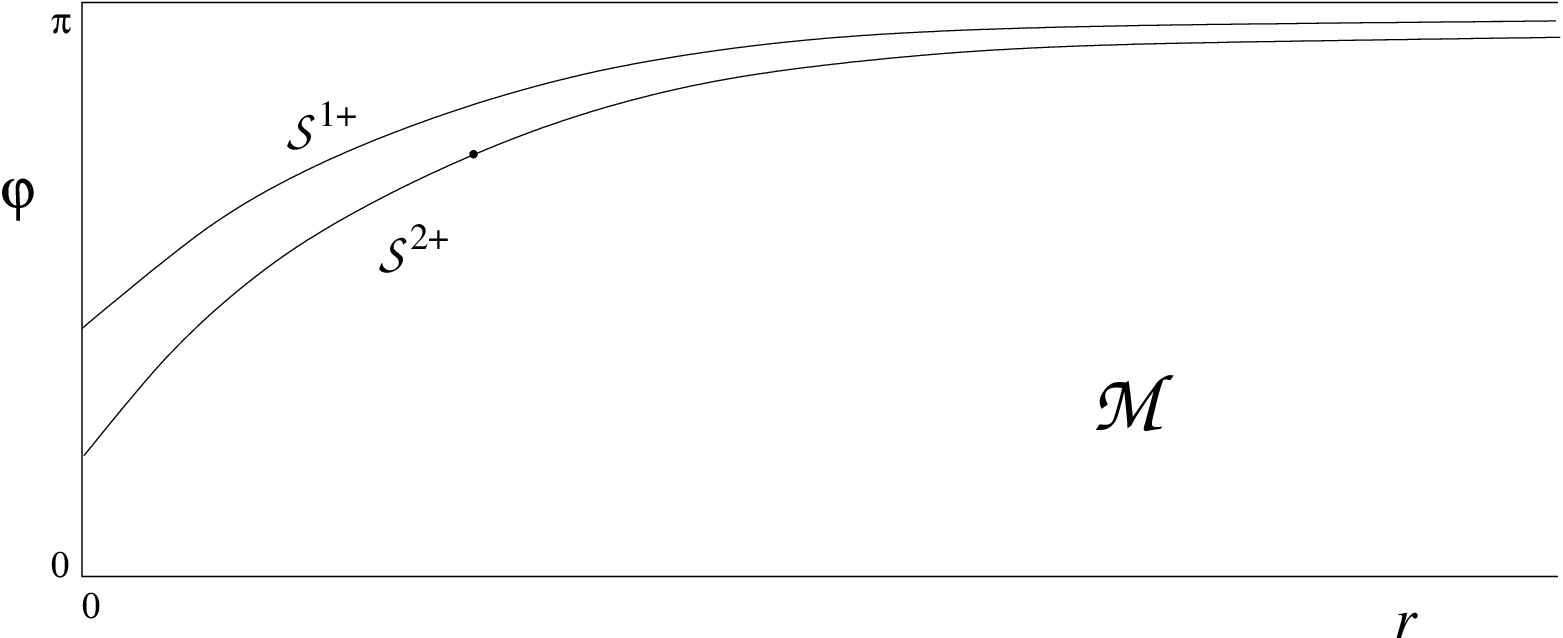} {5in} {The singularity lines $\si^{1+}$ and $\si^{2+}$ in
$\ps$.}

\sect{Hyperbolicity}
\label{sec-hyp}

In order to show that a given \bi\ is \erg\ we must first prove that
the associated dynamical \sy\ $(\ps, \ma, \mu)$ has a hyperbolic
structure. This means that for $\mu$-almost every point $\z \in \ps$,
there is a local stable and unstable manifold \cite[Defn.~6.1]{le2},
and the resulting foliations are absolutely continuous w.r.t.\
$\mu$. Sections 6--7 of \cite{le2} do precisely that, provided $\f$
satisfies five assumptions that we list as soon as we have given some
necessary notation.

\fig{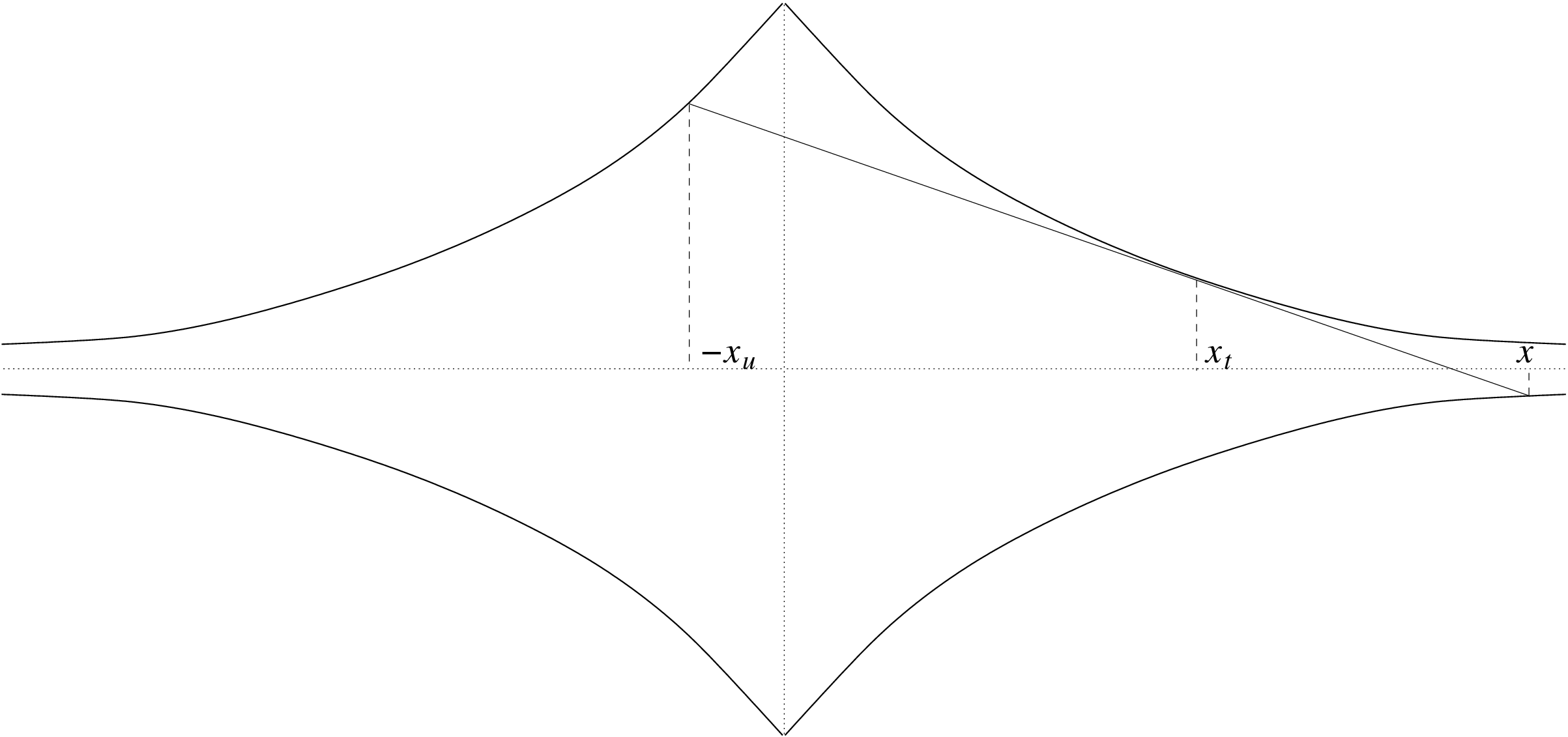} {5in} {The definition of $x_t$ and $x_u$.} 

In $\ta_4$, for $x > 0$, consider the straight line passing through
$(x,-\f(x))$ and tangent to $\partial \ta$ (i.e., to the part of
$\partial \ta_4$ that lies in the first quadrant). Denote by $x_t =
x_t(x)$ the abscissa of the tangency point. This is uniquely
determined by the equation 
\begin{equation}
	\frac{\f(x) + \f(x_t)} {x - x_t} = -\f'(x_t).
	\label{cond-xt}
\end{equation}
Then denote by $-x_u = -x_u(x) <0$ the abscissa of the point at which
this line intersects $\partial \ta_4$ in the second quadrant. This
number is given by the solution of
\begin{equation}
	\frac{\f(x) + \f(x_u)} {x + x_u} = -\f'(x_t).
	\label{cond-xu}
\end{equation}

In the sequel, for $f,g \ge 0$, we use the notation $f(x) \ll g(x)$ to
indicate that there is constant $C$ such that $f(x) \le C\, g(x)$, as
$x \to +\infty$; likewise for $\gg$. Also, $f(x) \sim g(x)$ means
that, when $x \to +\infty$, $f(x)/g(x)$ is bounded away from 0 and
$+\infty$; whereas $f(x) \simeq g(x)$ means that $f(x)/g(x) \to 1$.

The five assumptions read as follows:
$$
	\f''(x) \to 0;
	\eqno{\mathrm{(A1)}}
$$
$$
	|\f'(x_t)| \ll |\f'(x)|;
	\eqno{\mathrm{(A2)}}
$$
$$
	\frac{\f(x) \f''(x)} {(\f'(x))^2} \gg 1;
	\eqno{\mathrm{(A3)}}
$$
$$
	\frac{|\f'''(x)|} {\f''(x)} \ll 1.
	\eqno{\mathrm{(A4)}}
$$
$$
	|\f'(x)| \gg (\f(x))^\theta, \qquad \mbox{for some } \theta>0.
	\eqno{\mathrm{(A5)}}
$$

\skippar

We now prove a result that will make it simpler to check that our
\fn s verify (A1)-(A5).

\begin{lemma}
	For any positive $\f$ as in \emph{(H1)-(H3)},
	\begin{displaymath} 
		| \f^{(i)}(x) | \simeq \left[ \prod_{j=0}^{i-1}
		(p+j) \right] \upsilon(x) \, x^{-p-i}, \qquad i=1,2
	\end{displaymath}
	and
	\begin{displaymath} 
		| \f'''(x) | \ll \upsilon(x) \, x^{-p-3}.
	\end{displaymath}
	\label{lemma-simeq}
\end{lemma}

\proof This is a simple verification. Let us work out explicitly the
case $i=2$. Via (H3),
\begin{equation}
	\f''(x) = \upsilon''(x) x^{-p} - 2p \upsilon'(x) x^{-p-1} + p
	(p+1) \upsilon(x) x^{-p-2} \simeq p (p+1) \upsilon(x) x^{-p-2}.
\end{equation}
Recall that both $\f'$ and $\f''$ have definite signs. For $\f'''$ we
have only one asymptotic bound, as $\f'''(x)$ might vanish somewhere.
\qed

Now, checking (A3) and (A4) is immediate. For (A1) we apply (H2) to
$\f''(x) \sim \upsilon(x) x^{-p-2}$. For (A5) we may select any
$\theta > (p+1)/p$: then the assertion is given by applying (H2) to
$(\upsilon(x))^{\theta - 1} x^{-\theta p + p + 1}$.

Working on (A2) will be more complicated. We need a few lemmas.

\begin{lemma}
	Given any $c \in (0,1)$, then, \emph{uniformly} in $\xi \in 
	[c,1]$,
	\begin{displaymath}
		\frac{\upsilon(x)} {\upsilon(\xi x)} \to 1,
	\end{displaymath}
	as $x \to +\infty$.
	\label{lemma-tecn-a}
\end{lemma} 

\proof Assume for simplicity that $\xi < 1$ (otherwise the above ratio
is simply 1). Using Lagrange's Mean Value Theorem, there exists a
$\tilde{\xi} = \tilde{\xi}(x,\xi) \in (\xi,1)$ such that
\begin{equation}
	\log \upsilon(x) - \log \upsilon(\xi x) = \frac{\upsilon'
	(\tilde{\xi}x) } {\upsilon(\tilde{\xi}x) } \, \frac{ (1-\xi) x
	\tilde{\xi} } {\tilde{\xi}}.
\end{equation}
This vanishes via (H3) and the fact that $(1-\xi) / \tilde{\xi} <
1/c$.  
\qed

\begin{lemma}
	If $x_t(x)$ is defined by \emph{(\ref{cond-xt})}, then there 
	exists a constant $\bar{\xi}$ such that
	\begin{displaymath}
		\frac{x_t(x)} x \to \bar{\xi}.
	\end{displaymath}
	\label{lemma-tecn-b}
\end{lemma}

\proofof{Lemma \ref{lemma-tecn-b}} We plug (H1) in (\ref{cond-xt}) and
utilize Lemma \ref{lemma-simeq}: there is a \fn\ $\phi(x) \to 1$ such
that 
\begin{equation}
	\frac{\upsilon(x) x^{-p} + \upsilon(x_t) x_t^{-p}} {x - x_t} 
	= \phi(x) p \, \upsilon(x_t) x_t^{-p-1}.
	\label{tecn-b-10}
\end{equation}
Set $\xi := x_t / x \in (0,1)$. After some algebraic manipulations,
this equation can be rewritten as
\begin{equation}
	\frac{\upsilon(x)} {\upsilon(\xi x)} \, \xi^{p+1} + [1 +
	\phi(x) p] \, \xi - \phi(x) p = 0.
	\label{tecn-b-20}
\end{equation}
We know the genesis of (\ref{tecn-b-20}) and so we know that, at least
for $x$ sufficiently large, it has one and only one solution $\xi_x$
in $(0,1)$. We claim that $\lim_{x \to +\infty} \xi_x = \bar{\xi}$,
the latter being the (unique) solution of
\begin{equation}
	\xi^{p+1} + [1 + p] \, \xi - p = 0
	\label{tecn-b-30}
\end{equation}
in $(0,1)$. This would yield Lemma \ref{lemma-tecn-b}.

\begin{lemma}
	Assume that the real \fn\ $F_k(\xi)$ is smooth (in $\xi$) for
	$\xi \in \R$ and $k \in \R^{+}$. Suppose that one can find a
	$\bar{\xi}$ and one of its neighborhoods $U$ such that
	\begin{displaymath}
		\lim_{k \to +\infty} F_k(\bar{\xi}) = 0
	\end{displaymath}
	and
	\begin{displaymath}
		\liminf_{k \to +\infty} \inf_{\xi \in U} | F'_k(\xi) | 
		> 0.
	\end{displaymath}
	Then, for large $k$, $F_k(\xi) = 0$ has a unique solution
	$\xi_k$ in $U$, and 
	\begin{displaymath}
		\lim_{k \to +\infty} \xi_k = \bar{\xi}.
	\end{displaymath}
	\label{lemma-tecn-c}
\end{lemma}

We do not prove Lemma \ref{lemma-tecn-c} for its proof is obvious, and
we apply it immediately to our case, setting $F_x(\xi)$ to be l.h.s.\
of (\ref{tecn-b-20}). We utilize $x$ in lieu of $k$. 

We check the first hypothesis of Lemma \ref{lemma-tecn-c} by plugging
$\xi = \bar{\xi}$ in (\ref{tecn-b-20}) and using Lemma
\ref{lemma-tecn-a} and (\ref{tecn-b-30}). As for the second
hypothesis, taking the derivative of $F_x(\xi)$ w.r.t.\ $\xi$ we get,
after some rearrangement,
\begin{equation}
	F'_x(\xi) = \frac{\upsilon(x)} {\upsilon(\xi x)} \, \xi^p
	\left[ p+1 - \frac{ \upsilon'(\xi x) } { \upsilon(\xi x) } \,
	\xi x \right] + 1 + \phi(x) p.
	\label{tecn-b-40}
\end{equation}
Now choose an open interval $U$ such that $\bar{\xi} \in U \subset
(0,1)$. Via (H3), the only term in (\ref{tecn-b-40}) that is preceded
by a minus sign vanishes, as $x \to +\infty$. Taking into account the
asymptotics of all other terms (which are positive, anyway) we obtain,
for $x$ sufficiently large,
\begin{equation}
	\inf_{\xi \in U} F'_x(\xi) \ge 1 + p/2,
\end{equation}
which implies the second condition of Lemma \ref{lemma-tecn-c}, and
thus Lemma \ref{lemma-tecn-b}.
\qed

By Lemma \ref{lemma-simeq},
\begin{equation}
	\frac{ |\f'(x_t)| } { |\f'(x)| } \simeq \frac{ \upsilon(x_t)
	} { \upsilon(x) } \left( \frac{x_t}x \right)^{-p-1}.
	\label{preconcl-a2}
\end{equation}
Therefore, through Lemmas \ref{lemma-tecn-a} and \ref{lemma-tecn-b},
we find a constant $C>1$ such that
\begin{equation}
	|\f'(x_t)| \simeq  C |\f'(x)|,
	\label{concl-a2}
\end{equation}
which implies (A2).

\sect{Ergodicity}
\label{sec-erg}

If we look at \cite[Sec.~9]{le2} we see that there are two results
that are needed for the proof of the \erg ity that do not descend from
(A1)-(A5). One is eqn.~(9.2) and the other is Lemma 9.3. Proving the
first result is easy, with the techniques developed in the previous
section.

\begin{proposition}
	\begin{displaymath}
		|\f'(x_u)| \ll |\f'(x)|.
	\end{displaymath}
	\label{prop-xu}
\end{proposition}

\proof First of all, let us plug (\ref{concl-a2}) into
(\ref{cond-xu}):
\begin{equation}
	\frac{\f(x) + \f(x_u)} {x + x_u} \simeq C |\f'(x)|.
	\label{xu-10}
\end{equation}
Then we use (H1) as in the proof of Lemma \ref{lemma-tecn-b}.  Setting
$\xi := x_u/x$ and proceeding as in
(\ref{tecn-b-10})-(\ref{tecn-b-20}), we get
\begin{equation}
	\phi(x) p \, \xi^{p+1} + [\phi(x) p - 1] \, \xi^p -
	\frac{\upsilon(\xi x)} {\upsilon(x)} = 0,
	\label{xu-20}
\end{equation}
for some \fn\ $\phi(x) \to C$. Once again, we know that (\ref{xu-20})
has one and only one solution $\xi_x \in (0,1)$ and, once again, we
claim that $\lim_{x \to +\infty} \xi_x = \bar{\xi}$. This time
$\bar{\xi}$ is the solution of
\begin{equation}
	Cp \, \xi^{p+1} + [Cp - 1] \, \xi^p - 1 = 0
	\label{xu-30}
\end{equation}
in $(0,1)$. Via a relation totally analogous to (\ref{preconcl-a2}),
this would prove that $|\f'(x_u)| \simeq C_1 |\f'(x_u)|$, for some
$C_1 > 1$, whence Proposition \ref{prop-xu}.

The idea is to use Lemma \ref{lemma-tecn-c} one more time. Let
$F_x(\xi)$ be given by the l.h.s.\ of (\ref{xu-20}). Substituting $\xi
= \bar{\xi}$ in (\ref{xu-20}) gives the first condition of Lemma
\ref{lemma-tecn-c}, through Lemma \ref{lemma-tecn-a} and
(\ref{xu-30}). Furthermore,
\begin{eqnarray}
	F'_x(\xi) &=& \phi(x) p (p+1) \, \xi^p + [\phi(x) p - 1] p \,
	\xi^{p-1} - \frac{\upsilon'(\xi x) \xi x} {\upsilon(\xi x)}
	\, \frac{\upsilon(\xi x)} {\xi \upsilon(x)} = \nonumber \\
	&=& \phi(x) p \, \xi^p + \frac{p}{\xi} \left\{ \phi(x) p \,
	\xi^{p+1} + [\phi(x) p - 1] \, \xi^p \right\} - \sigma(x;\xi).
	\label{xu-40}
\end{eqnarray}
Here $\sigma(x;\xi)$, whose definition is implicit in the above,
vanishes as $x \to +\infty$ and $\xi$ remains away from 0 (due to
(H3) and Lemma \ref{lemma-tecn-a}).

Focussing our attention on the term inside the braces, we see that it
goes to 1, if we take $\xi = \bar{\xi}$---which is the solution of
(\ref{xu-30}). Therefore, $F'_x(\bar{\xi}) \to C p \, \bar{\xi}^p + p /
\bar{\xi}$.  Furthermore, as all the terms containing $x$ are well-behaved
when $x \to +\infty$, there is a neighborhood $U$ of $\bar{\xi}$ such that
\begin{equation}
	\inf_{\xi \in U} F'_x(\xi) \ge \frac12 \left( C p \, \bar{\xi}^p +
	\frac{p} {\bar{\xi}} \right),
\end{equation}
for $x$ large enough. This is the second hypothesis of Lemma
\ref{lemma-tecn-c}. Proposition \ref{prop-xu} is thus proved.
\qed

Lemma \ref{lemma-tecn3} that we are going to give in the second part
of Section \ref{sec-erg} substitutes Lemma 9.3 of \cite{le2}. Since we
present it out of context and the result is rather technical anyway,
we had better precede it with a little preamble.

\skippar

A point $(x,y) \in \uw$ can be parametrized in two ways: $\r$, the
arc-length coordinate along $\uw$, and $x$, the abscissa of the point
in the plane. The transformation between the two coordinates is given
by
\begin{equation}
	\r(x) := \int_0^x \sqrt{1 + (\f'(t))^2} \, dt.
	\label{r-x}
\end{equation}
Sometimes, like in the remainder, is convenient to use $x$. We do so
without worrying too much about a rigorous notation. For example we
will write $(x,\ph)$ to mean the point $(\r(x),\ph) \in \ps$.  With
this in mind, define
\begin{equation}
	h_2(x) := K_2 |\f'(x)|,
	\label{h2}
\end{equation}
for a certain constant $K_2$ provided by the \sy\ itself (the
subscript here has the sole purpose of keeping consistency of notation
with the corresponding objects in \cite{le2}). For $x$ large enough,
$h_2(x) < \pi/2$ and it makes sense to consider the line element $\w
:= \w_0 := (x,\ph_0) := (x,h_2(x))$. The associated unit vector points
more and more towards the cusp, as $x \to +\infty$; thus it will take
the material point more and more rebounds to ``come back from the
cusp''. In other words, setting $\w_i := (x_i, \ph_i) := \ma^i \w$ and
denoting by $m$ the first time (i.e., the first $\ma$-iteration) that
the \bi\ \tr y of $\w_i$ crosses the $y$-axis in $\ta_4$, we have $m =
m(x) \to +\infty$. The situation is illustrated in Fig.~\ref{figt}.

\fig{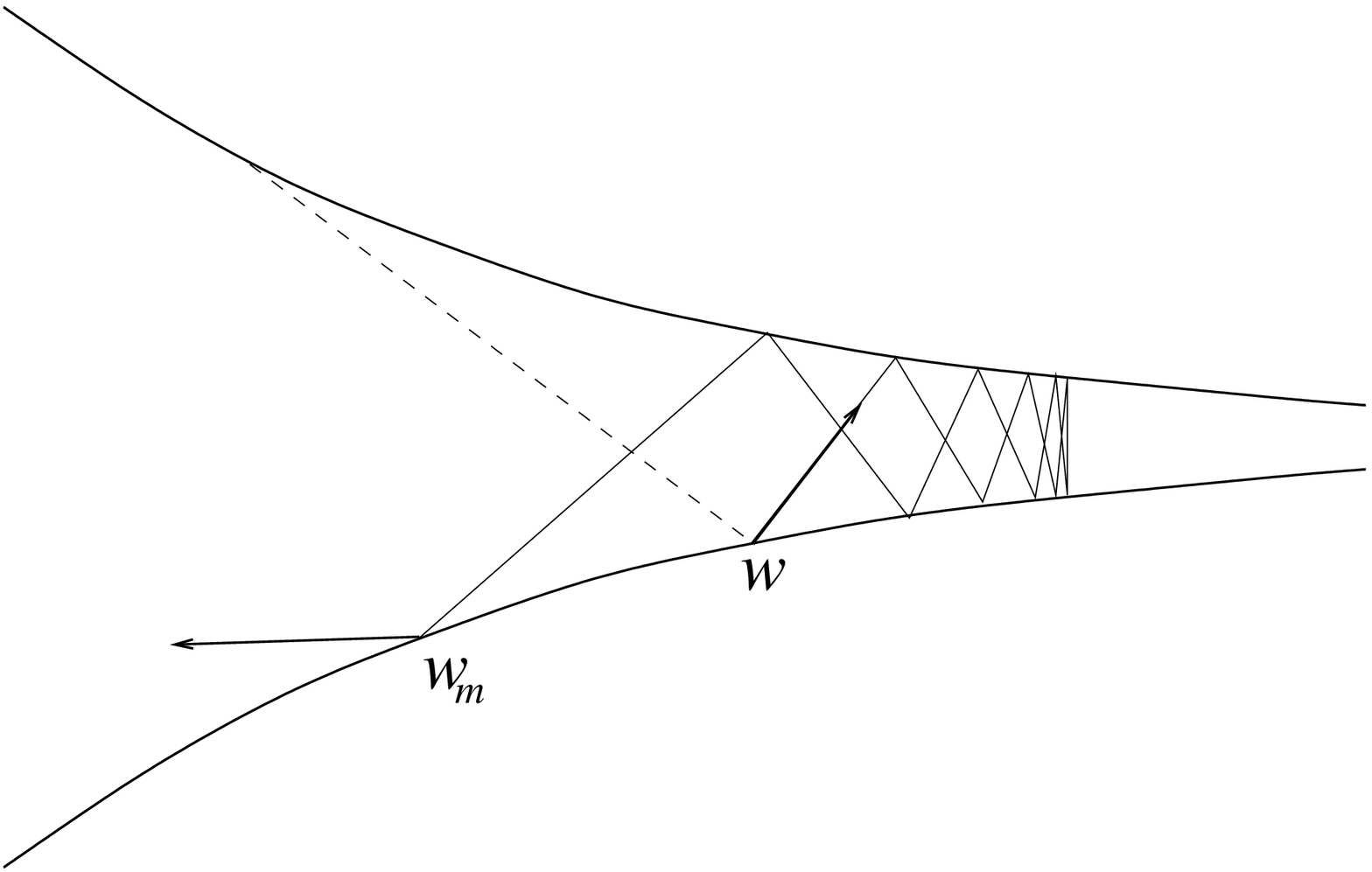} {4in} {For $\ph$ sufficiently small, the \tr y of $\w =
(\r,\ph)$ travels toward the cusp and comes back. The number $m$ of
collisions the material point makes against $\partial \ta_4$ before
crossing the $y$-axis tends to $\infty$ as $\ph \to 0$ and/or $\r \to
+\infty$.}

The recursive formula to determine $x_{n+1}$ is easily found out to be
\begin{equation}
	\frac{ \f(x_{n+1}) + \f(x_n) } { x_{n+1} - x_n } = \tan( \ph_n
	+ \alpha_n ), 
	\label{x-n+1}
\end{equation}
with $\alpha_n := \arctan |\f'(x_n)|$. It is not difficult either to
see that $\ph_{n+1} = \ph_n + \alpha_n + \alpha_{n+1}$ \cite{le2},
whence
\begin{equation}
	\ph_n = \ph_0 + \alpha_0 + 2 \sum_{i=1}^{n-1} \alpha_i +
	\alpha_n.
	\label{ph-n}
\end{equation}

Notice that all quantities ultimately depend on $x$. We are now ready
to state and prove our last result.

\begin{lemma}
	For $\f$ as in \emph{(H1)-(H3)}, there exists an increasing
	sequence $\{ \xi_n \}$ such that, for fixed $n$,
	\begin{displaymath}
		\lim_{x \to +\infty} \frac{x_n(x)} {x} = \xi_n.
	\end{displaymath}
	Furthermore
	\begin{displaymath}
		\lim_{n \to +\infty} \xi_n = +\infty.
	\end{displaymath}
	\label{lemma-tecn3}
\end{lemma}

\proof As concerns the first assertion, we prove it by induction. For
$n=0$ there is nothing to prove, as $x_0 = x$ (whence $\xi_0 = 1$). So
let us assume that the limit above exists for all $i = 1, ... ,n$, and
show that it exists for $n+1$, too.

As $n$ is fixed, all $\alpha_i(x)$, $i = 1, ... ,n$, tend to zero,
when $x \to +\infty$. And so does $\ph_n(x)$, from (\ref{ph-n}) and
the fact that $\ph_0(x) = h_2(x) = K_2 |\f'(x)|$---the latter coming
from (\ref{h2}). Therefore,
\begin{equation}
	\tan( \ph_n + \alpha_n ) \simeq \ph_n + \alpha_n \simeq K_2
	|\f'(x)| + |\f'(x)| + 2 \sum_{i=1}^n |\f'(x_i)|.
	\label{tecn3-20}
\end{equation}
Plugging (\ref{tecn3-20}) in (\ref{x-n+1}) we get, with the aid of
Lemma \ref{lemma-simeq},
\begin{equation}
	\frac{ \upsilon(x_{n+1}) x_{n+1}^{-p} + \upsilon(x_n) x_n^{-p}
	} { x_{n+1} - x_n } \simeq p \left[ (K_2+1) \upsilon(x)
	x^{-p-1} + 2 \sum_{i=1}^n \upsilon(x_i) x_i^{-p-1} \right].
	\label{tecn3-30}
\end{equation}
Divide both sides by $\upsilon(x) x^{-p-1}$:
\begin{equation}
	\frac{ \ds \frac{\upsilon(x_{n+1})} {\upsilon(x)} \left(
	\frac{x_{n+1}}x \right)^{-p} + \frac{\upsilon(x_n)}
	{\upsilon(x)} \left( \frac{x_n}x \right)^{-p} } { \ds \left(
	\frac{x_{n+1}}x \right) - \left( \frac{x_n}x \right) } =
	\phi(x) p \left[ K_2 + 1 + 2 \sum_{i=1}^n \frac{\upsilon(x_i)}
	{\upsilon(x)} \left( \frac{x_i}x \right)^{-p-1} \right],
	\label{tecn3-33}
\end{equation}
with $\phi(x) \to 1$ to replace the $\simeq$ sign.  Let us name
$\phi_1(x)$ the above r.h.s. For $i = 1, ... ,n$, by induction
hypothesis, $x_i/x \to \xi_i$ and, by Lemma \ref{lemma-tecn-a},
$\upsilon(x_i) / \upsilon(x) \to 1$. Therefore $\phi_1(x) \to
\tilde{\phi}_1$, for some $\tilde{\phi}_1 > 0$. Setting $\xi :=
x_{n+1}/x$ and rewriting (\ref{tecn3-33}) conveniently, we obtain
\begin{equation}
	\frac{\upsilon(\xi x)} {\upsilon(x)} \, \xi^{-p} - \phi_1(x)
	\xi = - \phi_1(x) \left( \frac{x_n}x \right) -
	\frac{\upsilon(x_n)} {\upsilon(x)} \left( \frac{x_n}x
	\right)^{-p} =: -\phi_2(x).  
	\label{tecn3-37}
\end{equation}
For the same reasons as before, $\phi_2(x) \to \tilde{\phi}_2 > 0$.
Multiplying by $\xi^p$, we get the equation
\begin{equation}
	F_x(\xi) := \phi_1(x) \, \xi^{p+1} - \phi_2(x) \, \xi^p -
	\frac{\upsilon(\xi x)} {\upsilon(x)} = 0.  
	\label{tecn3-38}
\end{equation}
It is now clear that we intend to apply again Lemma \ref{lemma-tecn-c}
and its associated arguments. Let $\xi_{n+1}$ be the (unique) solution
of
\begin{equation}
	\tilde{\phi}_1 \, \xi^{p+1} - \tilde{\phi}_2 \, \xi^p - 1 = 0
	\label{tecn3-39}
\end{equation}
in $(0,1)$. (Here $\xi_{n+1}$ takes on the role of $\bar{\xi}$ as in
the statement of Lemma \ref{lemma-tecn-c}.) In view of the asymptotics
of (\ref{tecn3-38}), one easily verifies that $F_x(\xi_{n+1}) \to 0$,
as $x \to +\infty$. Furthermore
\begin{eqnarray}
	F'_x(\xi) &=& \phi_1(x)(p+1) \, \xi^p - \phi_2(x) p \,
	\xi^{p-1} - \sigma(x;\xi) = \nonumber \\ 
	&=& \phi_1(x) \, \xi^p + \frac{p}{\xi} \left\{ \phi_1(x) 
	\, \xi^{p+1} - \phi_2(x) \, \xi^p \right\} - \sigma(x;\xi)
	\label{tecn3-42}
\end{eqnarray}
where $\sigma(x;\xi)$ is precisely the same as in (\ref{xu-40}), and
thus vanishes uniformly for $\xi \in [c,1]$. Also as in (\ref{xu-40}),
the term inside the braces goes to 1, when $\xi =
\xi_{n+1}$. Proceeding exactly as in the proof of Proposition
\ref{prop-xu}, we can take $U$ to be a sufficiently small neighborhood
of $\xi_{n+1}$ such that
\begin{equation}
	\inf_{\xi \in U} F'_x(\xi) \ge \frac12 \left( \tilde{\phi}_1 p
	\, \xi_{n+1}^p + \frac{p} {\xi_{n+1}} \right),
	\label{tecn3-44}
\end{equation}
for $x$ large enough. This allows us to apply Lemma \ref{lemma-tecn-c}
and conclude that $x_{n+1} / x = \xi \to \xi_{n+1}$. 

\skippar

The sequence $\{ \xi_n \}$ is increasing since $x_{n+1} > x_n$, at
least for $x$ sufficiently large (so that $\ph_n < \pi/2$ and the
material point moves to the right, after the $n$-th collision).

\skippar

As concerns the second assertion of the lemma, we proceed by
contradiction. Suppose that $\xi_n \nearrow \tilde{\xi} < +\infty$, as
$n \to +\infty$. Applying the first assertion to (\ref{tecn3-33}) one
obtains
\begin{equation}
	\frac{ \xi_{n+1}^{-p} + \xi_n^{-p} } { \xi_{n+1} - \xi_n }
	= p \left[ K_2 + 1 + 2 \sum_{i=1}^n \xi_i^{-p-1} \right].
	\label{tecn3-50}
\end{equation}
Then the r.h.s.\ of (\ref{tecn3-50}) grows asymptotically like $n$.
But the numerator of the l.h.s.\ converges, implying that $\xi_{n+1} -
\xi_n \sim n^{-1}$, which in turn contradicts the convergence of $\{
\xi_n \}$.  
\qed

\end{document}